\documentclass[aps,twocolumn,prb,superscriptaddress,showpacs,showkeys,floatfix]{revtex4}
\usepackage{epsfig}
\usepackage{makeidx} \makeindex
\usepackage{amsmath} \usepackage{amssymb} \usepackage{amsthm}
\usepackage{mathrsfs}
\DeclareMathAlphabet{\mathpzc}{OT1}{pzc}{m}{it}
\usepackage{bbm,eufrak}

\usepackage{color}

\begin{document}

\title{Optimal tuning of solid-state quantum gates: A universal two-qubit gate}

\begin{abstract}
We present a general route to reduce  
inhomogeneous broadening in nanodevices due to $1/f$ noise.  
We apply this method to a universal two-qubit gate and demonstrate that 
for selected {\em optimal couplings}, a high-efficient  gate
can be implemented even in the presence of $1/f$ noise. 
Entanglement degradation due to interplay of $1/f$ and quantum noise is
quantified via the concurrence. 
A charge-phase $\sqrt{{\rm i-SWAP}}$ gate
for spectra extrapolated from single qubit experiments is analyzed.
\end{abstract}

\author{E. Paladino} \affiliation{Dipartimento di Metodologie Fisiche e Chimiche (DMFCI), 
Universit\`a  di Catania. Viale A. Doria 6, 95125 Catania (Italy) 
\& MATIS  CNR - INFM, Catania}
\author{A. Mastellone}
\affiliation{C.I.R.A. Centro Italiano Ricerche Aerospaziali -
Via Maiorise snc - 81043 Capua, Caserta (Italy)}
\affiliation{Dipartimento di Metodologie Fisiche e 
Chimiche (DMFCI), Universit\`a di Catania. Viale A. Doria 6, 95125 Catania (Italy) 
\&  MATIS  CNR - INFM, Catania}
\author{A. D'Arrigo} \affiliation{Dipartimento di Metodologie Fisiche e 
Chimiche (DMFCI), 
Universit\`a  di Catania. Viale A. Doria 6, 95125 Catania (Italy) 
\& MATIS CNR - INFM, Catania}
\author{G. Falci} \affiliation{Dipartimento di Metodologie Fisiche e 
Chimiche (DMFCI), Universit\`a  di Catania. Viale A. Doria 6, 95125 Catania (Italy) 
\& MATIS  CNR - INFM, Catania}

\email[email: ]{epaladino@dmfci.unict.it}


\maketitle

Intense research on solid state nanodevices during the last decade
lead to observation of  fundamental quantum phenomena 
at the nanoscale. Combining quantum coherence with 
the existing integrated-circuit fabrication technology makes
nanocircuits very promising for quantum computing. In particular,
a variety of high-fidelity single qubit gates  
on a superconducting platform
are nowadays  available~\cite{single-super,kn:vion,chiarello}. 
Controlled generation of entangled states and preserving
quantum correlations represent a timely and critical issue. 
Identification of strategies to counteract 
physical processes detrimental to quantum coherent behavior 
is a fundamental step towards this goal.

Noise with $1/f$ spectrum is ubiquitous in 
nanodevices and represents a serious limitation 
for their coherence properties~\cite{single-super,kn:vion,chiarello,oneoverf,nak-eco,PRL02}.
Because of low frequency fluctuations of the device eigenenergies,
the average (implied by quantum measurement) of signals 
occurring at slightly different frequencies, is defocused.
This effect, analogous to inhomogeneous broadening~\cite{slichter},
is a signature of $1/f$ noise, whatever its microscopic origin. 
It results in a typical 
algebraic decay of coherent
oscillations~\cite{PRL05}, as observed in single-qubit gates of 
various design and materials~\cite{kn:vion,chiarello}.

One way to defeat inhomogeneities is exploiting NMR decoupling procedures~\cite{reviewControl}.
For instance, echo-type protocols may considerably reduce
defocusing in
single qubit rotations~\cite{nak-eco,kn:vion,fluxNEC}, but they also limit  
external control on quantum state processing.
Moreover, extension to multi-qubit gates may require a large fraction of quantum operations  
devoted to effective decoupling procedures, limiting scalability.
A simpler but effective strategy is to 
operate the device at ``optimal points'', characterized by minimal sensitivity of the relevant
splittings to variations of the control parameters. 
Partial reduction of defocusing of single-qubit operations by operating at such
working point has been successfully demonstrated~\cite{single-super,kn:vion,fluxNEC}.

In this Brief Report, we extend this scheme and introduce a general route to reduce 
inhomogeneous broadening effects in nanodevices. 
The strategy we propose  exploits tunability to counteract the effect 
of noise channels opened up by control parameters
themselves. 
Beside the interest for quantum gates  
engineering, on a more fundamental level our analysis
is relevant for optimizing  fault-tolerant architectures~\cite{knill},
showing that the influence of $1/f$ 
fluctuations in the solid state can be limited by
exploiting the band structure of coupled nanodevices.

We apply this method to a universal two-qubit gate involving an entanglement
operation on two quantum bits,
a necessary step toward the
construction of a scalable quantum computer~\cite{Nielsen}.
We consider a $\sqrt{\mathrm{i-SWAP}}$ gate based on a fixed coupling scheme and
show that, for selected {\em optimal couplings}, it can be accurately operated
even in the presence of 
$1/f$ and high-frequency noise.
As a relevant example, we analyze a realistic
charge-phase two-qubit gate~\cite{nguyen}.

{\em Optimal tuning --}
We denote the device Hamiltonian with ${\cal H}_0$, and its eigenvalues
as $\omega_i$. 
Fluctuations of the control parameters due to external environments 
are modeled by 
$\mathcal{H}_\mathrm{I} = -\frac{1}{2} \sum_\alpha \hat Q_\alpha \otimes \hat{X}_{\alpha}$. 
Here $\{\hat X_\alpha\}$ are collective 
environmental variables coupled to the nanodevice operators, $\hat Q_\alpha $.
Typically, 
power spectra of $\hat X_\alpha$  
display a $1/f$ low-frequency  behavior  up 
to some cut-off, followed by a white or ohmic flank~\cite{kn:vion,nak-spectrum}. 
Defocusing is due to low-frequency fluctuations whereas noise at high-frequencies 
(of the order of the eigen-splittings) 
is responsible for system-bath inelastic energy exchanges. 
Acting on different time scales, these two processes can be considered separately.  
Upon extending 
the  multi-stage elimination approach \cite{PRL05}, we separate the effects of low- 
and high-frequency components 
of the noise by putting, 
$\hat{X}_\alpha \to X_\alpha(t) + \hat X_\alpha^f$.
Here  $ \{X_\alpha(t)\} =  \mathbf{X}(t)$ are classical stochastic variables 
describing low-frequency
($1/f$) noise, and can be treated  
in the adiabatic approximation (adiabatic noise).
Instead, high-frequency 
fluctuations $\hat  X_\alpha^f$ 
are modeled by a Markovian bath and mainly determine 
spontaneous decay (quantum noise).
Nanodevice populations relax  due to quantum noise 
($T_1$-type times),  which also leads to secular dephasing 
($T_2^*= 2 \, T_1$-type). 
Defocusing due to low-frequency noise determines further decay of 
the coherences.

In the adiabatic and longitudinal approximation~\cite{PRL05} the system evolution  
is related to instantaneous eigenvalues, $\omega_i(\mathbf{X}(t))$, which
depend on the noise realization. 
The leading effect of low-frequency fluctuations in repeated measurements 
is given within the Static Path Approximation (SPA). It amounts to replace, 
in the path-integral average over $\mathbf{X}(t)$ realizations,
$X_\alpha(t)$
with statistically distributed values $X_\alpha(0) \equiv X_\alpha$ at each repetition
of the measurement protocol. 
As a consequence, level splittings $ \omega_{ij}(\mathbf{X})$ are 
random variables, with standard deviation
$\Sigma_{ij}=\sqrt{\langle\delta \omega^2_{ij}\rangle-
\langle\delta \omega_{ij}\rangle^2}$, where $\delta \omega_{ij} = \omega_{ij}(\mathbf{X})-
\omega_{ij}$. The splittings enter the evolution of the reduced density matrix 
coherences in the eigenbasis of
${\cal H}_0$ as 
$\rho_{ij}(t) \approx \rho_{ij}(0) \int d \mathbf{X}  
P(\mathbf{X}) \exp[- i \omega_{ij}(\mathbf{X}) t]$.
The probability density can be taken of Gaussian
form $P(X_\alpha) = \exp[- X_\alpha^2/2 \Sigma_{X_\alpha}^2]/\sqrt{2 \pi} \Sigma_{X_\alpha}$ in
relevant cases~\cite{PRL05}.  
Optimal tuning of control parameters is achieved when
the variance of the relevant instantaneous
splittings, $\Sigma_{ij}$, is minimized. 
It results in the reduction of $\rho_{ij}(t)$ decay due to defocusing processes.

If $ \omega_{ij}(\mathbf{X})$  is monotonic 
in a region $|X_\alpha| \leq 3 \Sigma_{X_\alpha}$, 
$\Sigma_{ij}^2 \approx \sum_\alpha \left [\frac{\partial \omega_{ij}}{\partial
X_\alpha}|_{X_\alpha=0}\right ]^2 \Sigma_{X_\alpha}^2$.
The variance attains a minimum for vanishing differential dispersion.
This is the case of single-qubit optimal points.
For the charge-phase two-port architecture, control is 
 via gate voltage, 
$q_{x}\equiv C_g V_g/(2e)$, and magnetic-flux dependent phase $\delta$,
entering the Josephson energy $E_{J} = E_{J}^0 \cos \delta$. 
Thus $X_\alpha \to \Delta q_{x}, \Delta E_J $ and
the optimal point, $q_x=1/2 \, , \delta=0$, 
is at the a saddle point of the energy bands~\cite{kn:vion}.
When bands are non-monotonic in the control parameters, 
minimization of defocusing necessarily requires their 
tuning  to values depending on the noise variances.
We illustrate potentialities of this  result for a universal two-qubit gate.

{\em $\sqrt{\rm{i-SWAP}}$ gate --}
A simple way to realize two-qubit entanglement is via a fixed, 
capacitive or inductive, coupling scheme~\cite{coupled-th-fixed}. 
Fast two-qubit operations and  coupling switching on/off 
are achieved by individual-qubit control.
Fixed coupling has been used to demonstrate
two-qubit logic gates~$^{\rm{1(e)},}$~\cite{coupled-exp-fix} and
high-fidelity Bell states generation in capacitive coupled phase qubits~\cite{Steffen}. 

The interaction is effectively switched on 
by tuning the single-qubit energy spacing 
to  mutual resonance. The building-block Hamiltonian reads
$\mathcal{H}_0 = 
 -\frac{\Omega}{2} \, \sigma_3^{(1)} \otimes \mathbbm{1}^{(2)}
 -\frac{\Omega}{2}  \, \mathbbm{1}^{(1)} \otimes \sigma_3^{(2)}
+ \frac{\omega_c}{2} \, \sigma_1^{(1)} \otimes \sigma_1^{(2)} $ ($\hbar=1$,
$\sigma_3^{(\alpha)}| \pm \rangle= \pm  | \pm \rangle $).
The Hilbert space factorizes in a "SWAP-subspace", spanned by
the eigenstates of ${\cal H}_0$ 
$\{ |1 \rangle \, = \,\frac{1}{\sqrt{2}}\, (- |+- \rangle  \, + \,| -+ \rangle )$ and 
$|2 \rangle \,= \, \frac{1}{\sqrt{2}} \,( | +- \rangle \, + \, |-+ \rangle ) \}$ 
(eigenvalues $\omega_{{}^{1}_{2}} = \mp \omega_c/2$),
and in a "Z-subspace" generated by the eigenstates 
$\{ |0 \rangle \, = \, - \sin \frac{\varphi}{2}   \,| ++ \rangle 
\, + \, \cos \frac{\varphi}{2}  \, | -- \rangle$ and 
$|3 \rangle \, = \,\cos \frac{\varphi}{2} \,  | ++ \rangle 
\, + \, \sin \frac{\varphi}{2}\, | -- \rangle\}$, where
$\tan \varphi = - \omega_c/(2 \Omega)$
(eigenvalues $\omega_{{}^{0}_{3}}= \mp  \sqrt{\Omega^2 + (\omega_c/2)^2}$).
We focus on the $\sqrt{{\rm i-SWAP}}$ operation 
$| +- \rangle \to |\psi_e \rangle = [| +- \rangle - i | -+ \rangle]/\sqrt{2}$
which generates  by free evolution an entangled state
at $t_E = \pi/2 \omega_c$.
We consider the general case where 
$\mathcal{H}_\mathrm{I} = -\frac{1}{2} \left [ \hat{x}_{1} 
\sigma_1^{(1)} +  \hat{z}_{1} \sigma_3^{(1)} \right ]
\otimes \mathbbm{1}_2
-
\frac{1}{2} \mathbbm{1}_1 \otimes \left [ 
\hat{x}_{2} \sigma_1^{(2)} + \hat{z}_{2} \sigma_3^{(2)} \right ]$. 
Since transverse ($\hat{x}_{i}$)
and longitudinal ($\hat{z}_{i}$) fluctuations usually have different physical origin, 
we assume they are independent~\cite{NJP-special}. 
For capacitive coupled charge-phase qubits, polarization fluctuations cause transverse noise 
$x_i \propto  4  E_{C,i} \Delta q_{x,i}$ ($ E_{C,i}$ qubit $i$ charging energy),
phase fluctuations lead to longitudinal noise
$z_i \propto \Delta E_{J,i} $~\cite{quantroswap}.
Both low-frequency 
fluctuations induce a stochastic effective SWAP
 splitting, $\omega_{21}(\{x_i,z_i\})$, 
  the relevant scale for the gate operation. It can be
obtained from $\mathcal{H}_0 + \mathcal{H}_\mathrm{I}$
treating perturbatively the stochastic fields in $\mathcal{H}_\mathrm{I}$,   
\begin{eqnarray}
&& \!\!\!\!\!\!\!\!\! \omega_{21}(x_1,x_2,z_1,z_2) \approx \omega_c -\frac{\omega_c}{2 \Omega^2}  (x_1^2+x_2^2) 
+ \frac{1}{2\omega_c}(z_1-z_2)^2 
	\nonumber \\
 &+& \!\! \frac{\omega_c}{2 \Omega^3}  (x_1^2 +x_2^2)(z_1+z_2) +  
\frac{1}{2\omega_c\Omega}  (x_1^2 - x_2^2)(z_1 - z_2)
 \label{eq:splittingSWAP} \\
 &+&  \!\!   \frac{\omega_c}{8 \Omega^4} 
(1+  \frac{\omega_c^2}{\Omega^2})
(x_1^4+ 6 x_1^2 x_2^2 + x_2^4) + \frac{1}{8\omega_c \Omega^2}(x_1^2-x_2^2)^2 . \nonumber 
\end{eqnarray}
\begin{figure}[t]
\centering
\resizebox{42mm}{34mm}{\includegraphics{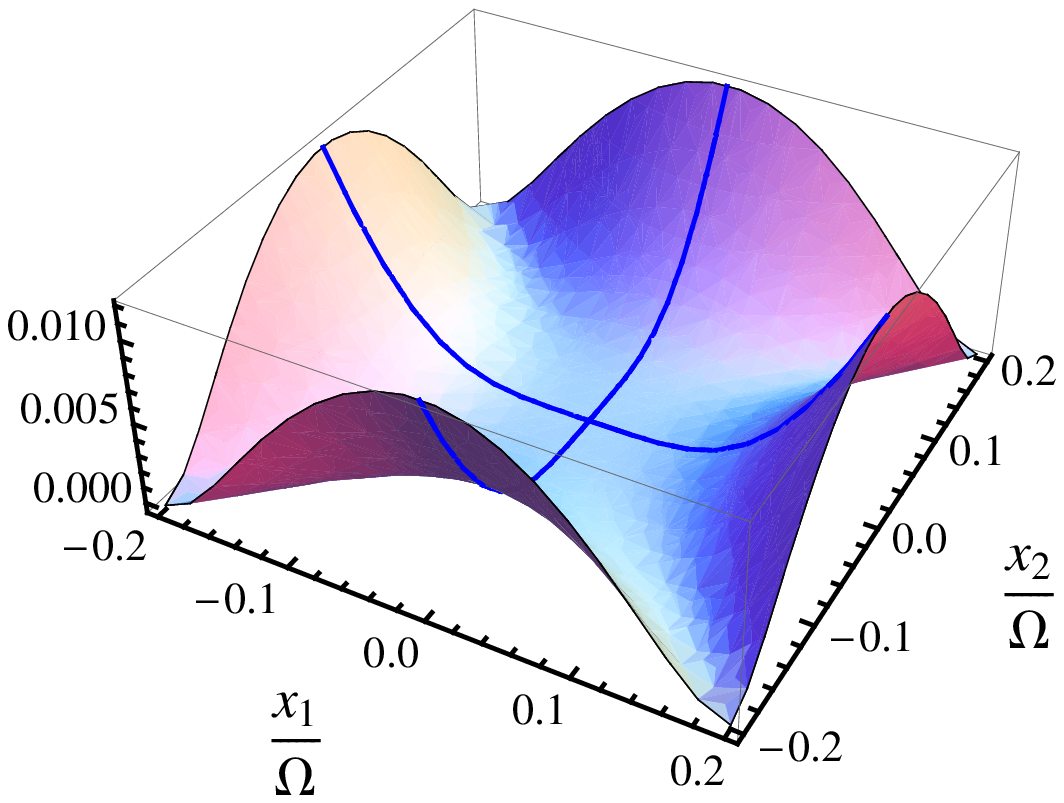}}
\resizebox{43mm}{27mm}{\includegraphics{paladinoFIG1b}}
\resizebox{42mm}{34mm}{\includegraphics{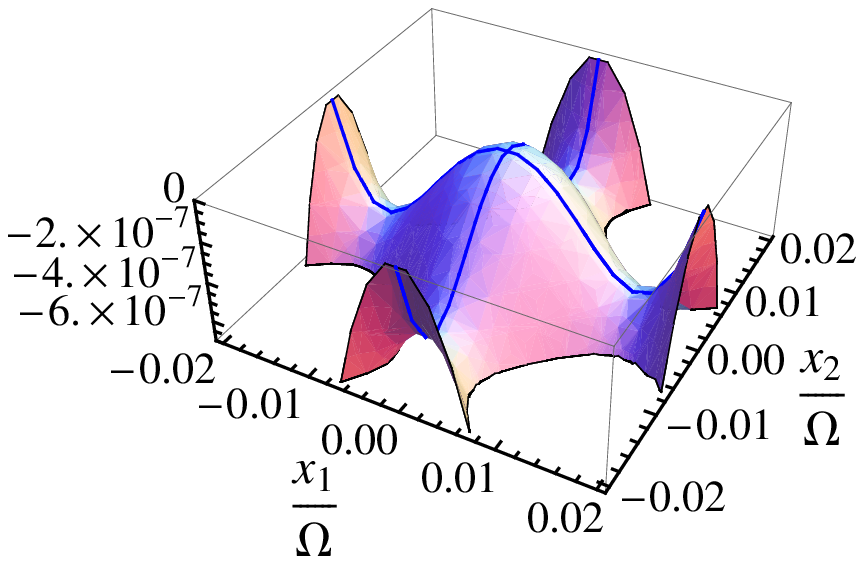}}
\resizebox{43mm}{27mm}{\includegraphics{paladinoFIG1d}}
\caption{(Color online) Dispersion of the SWAP and Z splittings for $\omega_c/\Omega=0.01$. 
Left: $\delta \omega_{21}/\Omega$ from numerical diagonalization of 
$\mathcal{H}_0 + \mathcal{H}_\mathrm{I}$ (top). 
Zoom around the origin highlights the interplay of $2^{nd}$ and $4^{th}$ order
terms, barrier height $\propto \omega_c^3$ (bottom).
Right: Comparative behavior of dispersions in the two subspaces.
Top: SWAP exact splitting  blue (gray) line, expansion (\ref{eq:splittingSWAP}) for 
$x_2=0$, $y_i=0$  dashed blue (gray), $2^{nd}$ order expansion (dotted), $Z$ splitting (thick gray)
$ \delta \omega_{30} 
\simeq   - \frac{\cos \varphi}{2}
\left \{ (y_1+y_2)  +  
\left [1 + \frac{1}{2}\left (\frac{\omega_c}{\Omega} \right )^2 \right  ]\frac{x_1^2+x_2^2}{\Omega}
\right\} $,
and single qubit dispersion  (circles).
Bottom: Longitudinal dispersions.
The $Z$-subspace (light gray) is much more sensitive both to transverse and longitudinal variations.}
\label{fig-bands}
\end{figure}
A key feature is that $\omega_{21}$ is non-monotonic in the small coupling  $\omega_c \ll \Omega$. 
This is due to a selection rule for longitudinal fluctuations. They only mix states inside each -
SWAP or Z - subspace, while $x_i$-fluctuations mix the two subspaces.
For instance, second order transverse corrections to $\omega_1$ are 
$\sum_{i \neq 1,2}  | \langle 1 | \mathcal{H}_\mathrm{I} |i \rangle|^2 /(\omega_1- \omega_i) \propto \omega_c$, 
whereas longitudinal ones  vary as 
$\omega_c^{-1} \propto |\langle 1 | \mathcal{H}_\mathrm{I} |2 \rangle |^2/(\omega_1-\omega_2)$.
Non-monotonicity in $\omega_c$ results in
a competition between $2^{nd}$ and $4^{th}$ order $x_i$-terms in Eq.(\ref{eq:splittingSWAP}) and in
non-monotonic band structure,  Figure 1 (left). 
Because of this subtle feature, 
identification of the best operating condition  
necessarily requires consideration of the noise characteristics.
Indeed an {\em optimal coupling} can be found 
which minimizes the SWAP variance 
\begin{equation}
\Sigma_{21}^2 \approx \frac{1}{\omega_c^2} 
 \left \{  \left (\frac{\Sigma_x}{\Omega} \right )^4 
\left [ (\Sigma_x^2 - \omega_c^2)^2 + \Sigma_x^4 
+ \Sigma_{z_2}^2 \Omega^2 \right ] + \frac{\Sigma_{z_2}^4}{2}   \right \} \,
\label{eq:variance}
\end{equation}
\begin{figure}[t]
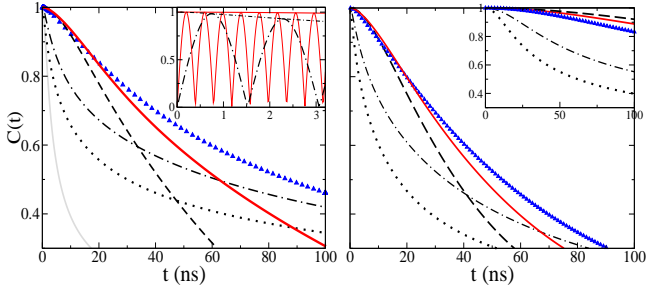

\centering
\resizebox{44mm}{38mm}{\includegraphics{paladinoFIG2a}}
\resizebox{40mm}{37.7mm}{\includegraphics{paladinoFIG2b}}
\caption{(Color online) Envelope of the concurrence in the SPA, $\Omega = 10^{11}$~rad/s. 
To elucidate the significance of the optimal coupling scheme, we consider  $1/f$ 
amplitudes larger than expected from single qubit measurements~\cite{quantroswap}.
Left: Effect of transverse noise with  $\Sigma_x/\Omega = 0.08$
and $\omega_c/\Omega= 0.01, 0.02,0.06,0.08,0.1$ (dotted, dot-dashed, triangles, thick (gray) red, dashed).
Light gray is the single qubit coherence, $|\rho_{+-}(t)| = [1+ (\Sigma_x^2t/\Omega^2)]^{-1/4}$
\cite{PRL05}. Inset: $C(t)$ and its envelope
for  $\omega_c/\Omega= 0.02,0.08$. For optimal coupling $\tilde \omega_c \approx \Sigma_x$, 
at $3$ns already $8$-SWAP  cycles occurred.
Right: Effect of transverse plus longitudinal noise
on qubit 2, $\Sigma_{z_2}/\Omega = 2.5 \times 10^{-3}$. Inset: 
Effect of longitudinal noise,
$\omega_c$ values as in left panel. 
}
\label{fig-SPA}
\end{figure}
where we assumed equal transverse variances,
$\Sigma_{x_i}= \Sigma_x$
and $\Sigma_{z_1} \ll \Sigma_{z_2}$, which mimic
typical experimental conditions~\cite{quantroswap}.
Eqs. (\ref{eq:splittingSWAP})~-~(\ref{eq:variance})
highlight the general result of this Brief Report.
While higher stability with respect 
to {\em longitudinal} fluctuations is
attained by larger couplings, minimization of the
detrimental {\em transverse} low-frequency noise 
components is obtained by tuning the  
coupling to an {\em optimal value} $\tilde\omega_c$. 
For $\Sigma_{z_2} \ll \Sigma_x$, 
this is the transverse noise variance, 
$ \tilde \omega_c \approx 2^{1/4}\Sigma_x$. 
It can be estimated by independent measurement 
of the amplitude of the $1/f$ transverse noise
on the uncoupled qubits,  
$S_x^{1/f} = \pi \Sigma_x^2  
[\ln(\gamma_M/\gamma_m) \, \omega]^{-1}$,
$\Sigma_x^2 = \int_{0}^\infty d \omega/\pi S_x^{1/f}(\omega)$ 
(low and high frequency cut-offs $\gamma_m$ and $\gamma_M$).
Note the higher stability of the SWAP-splitting 
compared  to the
qubits Larmor frequency and the Z-splitting,  Figure 1. 

Working at the optimal coupling 
minimizes defocusing and
guarantees excellent performance of 
the $\sqrt{{\rm i-SWAP}}$ operation. 
As a unambiguous test of entanglement generation and 
its degradation due to noise, 
we calculated the  concurrence during the gate 
operation~\cite{concurrence}. 
If the system is initialized in  $| +- \rangle$, 
the $4 \times 4$ density matrix
in the computational basis is non-vanishing 
only along the diagonal and anti-diagonal at any time (X-states)
and the concurrence takes a simple form~\cite{Xform}.

We first consider the effect of low-frequency fluctuations.
In the adiabatic approximation, 
the concurrence simplifies to 
$C(t)= 2 \,|{\rm Im} \{\rho_{12}(t)\}|$.
With the SWAP-splitting expansion (\ref{eq:splittingSWAP})  
(including in the $3^{rd}$ and $4^{th}$ order the terms   
$\propto \omega_c^{-1}$) in the SPA we obtain
\begin{equation}
\rho_{12}(t) = \rho_{12}(0)  \,
\frac{ \Omega }{2 \Sigma_x^2} \sqrt{\frac{2 i
\omega_c}{\pi  t}} \, 
e^{i \omega_c t+ h(t)} \, K_0[ h(t)] 
\label{eq:SPA-SWAP}
\end{equation}
where 
$h(t)= (\Sigma_{z_1}^2 +\Sigma_{z_2}^2 + i \omega_c/t) \, 
(\Omega^2/\Sigma_x^2 +i \omega_c t)^2/(4 \Omega^2)$,
and $K_0[h]$  is the K-Bessel function of order zero~\cite{abramowitz}.
By increasing the coupling to match the optimal condition two 
goals are simultaneously achieved: 
minimization of initial defocusing 
and fast two-qubit gate (Figure \ref{fig-SPA}).
The first SWAP error  takes remarkable values 
$\varepsilon =  1 - \langle \psi_e | \rho(t_E) | \psi_e \rangle \approx 10^{-3} -  10^{-4}$, 
for  $\omega_c \approx \Sigma_x \leq 0.05 \Omega$ (numerical simulations).
This is an interesting effect, considering that single qubit coherence times would be
rather small  at the same $1/f$ amplitudes, $T_2 \approx 5$~ns. 
\begin{figure}[t]
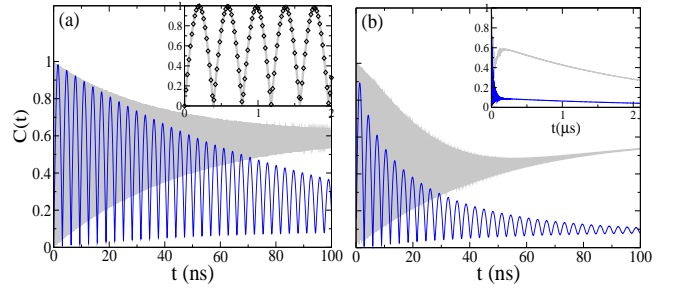

\centering
\resizebox{44mm}{38mm}{\includegraphics{paladinoFIG3a}}
\resizebox{40mm}{37.7mm}{\includegraphics{paladinoFIG3b}}
\caption{(Color online) Concurrence for $\omega_c/\Omega= 0.01$ blue (gray) line and for optimal coupling  
$\tilde \omega_c/\Omega=0.08$ (light gray).
(a) Effect of high-frequency noise, $S_{x_i}(\omega) \approx 8 \times 10^5$~s$^{-1}$, 
$S_{z_2}(\omega) 
\approx 4 \times 10^7$~s$^{-1} \gg S_{z_1}(\omega_c=0.08 \Omega) \approx 6 \times 10^3$~s$^{-1}$
~\cite{PhysicaE}.
Inset: At short times  
$C(t) \approx 2 |{\rm Im}\{\rho_{12}(t)\}|$ (diamonds).
(b) Effect of $1/f$ noise (parameters as in Fig.~\ref{fig-SPA}) 
and white quantum noise. Inset: Asymptotic behavior. 
Results are minimally modified considering the dynamics of 
fluctuators generating $1/f$ transverse (longitudinal) noise 
in $\gamma_m = 1$~s$^{-1}$, $\gamma_M=10^6 \,(10^{10})$~s$^{-1}$ 
(numerical solution of the stochastic Schr\"odinger equation). 
}
\label{concurrence2}
\end{figure}
The optimal coupling scheme is effective against large amplitude
$1/f$~-~noise even in the presence of  high-frequency fluctuations.
Within the  secular approximation,  quantum noise 
leads to additional exponential decay of SWAP-coherence,
$\widetilde \rho_{12}(t) = \rho_{12}(t) \, \exp{\{- \widetilde \Gamma_{12} t\}}$.
Thermal relaxation processes ($k_B T \ll \Omega$) populate levels $i= 0 - 2$ and
the concurrence reads
$C(t) \approx \sqrt{ ( \rho_{11} - \rho_{22} )^2 + 2 ({\rm Im }[\widetilde \rho_{12}])^2 } 
- |\sin \varphi| \rho_{00}$.
The SWAP decay rate is related to escape rates from levels $1$ and $2$,
$\widetilde \Gamma_{12} \approx  \frac{\Gamma_1^{\rm e} + \Gamma_2^{\rm e}}{2}$,
where
$\Gamma_1^{\rm e} = \Gamma_{10} + \Gamma_{12}$,  
$\Gamma_2^{\rm e} = \Gamma_{20} + \Gamma_{21}$ and 
$\Gamma_{i0} \propto [ S_{x_1}(\omega_{i0}) +  S_{x_2}(\omega_{i0})]$, 
$\Gamma_{21} \propto [S_{z_1}(\omega_{21}) + S_{z_2}(\omega_{21})]$. 

These rates enter the populations 
in the  combinations
$ 
\Gamma_{\pm} = - (\Gamma_1^{\rm e} + \Gamma_2^{\rm e})/2 
\pm \, \sqrt{(\Gamma_1^{\rm e} - \Gamma_2^{\rm e})^2 
+ 4 \Gamma_{12} \Gamma_{21}}/2 
$~\cite{PhysicaE}.
The  SWAP-coherence rules the relevant short-time behavior,
$|\rho_{12}| \propto 1- \widetilde \Gamma_{12} t$ (or $|\rho_{12}|\propto 1 - (\Sigma_{21} t)^2/2$)
depending on the most relevant
quantum (or adiabatic) noise component.
Exponential long-time decay is due to the strongest longitudinal or transverse
quantum noise, Figure \ref{concurrence2}. 
The finite asymptotic value 
reflects the entangled thermalized state
(no ``entanglement sudden death'' occurs~\cite{Xform}).
\begin{figure}[t]
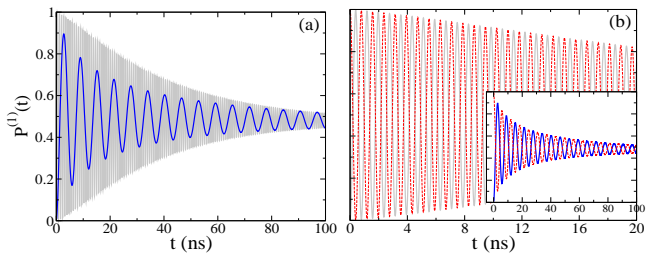

\centering
\resizebox{44mm}{33mm}{\includegraphics{paladinoFIG4a}}
\resizebox{40mm}{32.7mm}{\includegraphics{paladinoFIG4b}}
\caption{(Color online) Qubit $1$ switching probability 
to $\vert - \rangle$,  $P^{(1)}(t)$ and  probability $P^{(2)}(t)$
to find  qubit $2$ in the initial state $\vert - \rangle$ in the presence of  $1/f$ 
and white noise for $\omega_c/\Omega= 0.01$ blue (gray) line and for optimal coupling
$\tilde \omega_c= \Sigma_x = 0.08 \Omega$ (light gray).
(a) 
$P^{(1)}$: 
Exponential short-time limit at $\tilde \omega_c$,  algebraic otherwise.
(b) $P^{(1)}$ and $P^{(2)}$ [dashed (gray) red]  anti-phase oscillations for
$\tilde \omega_c$ (main), $ \omega_c/\Omega= 0.01$ (inset).  
}
\label{switching}
\end{figure}
The above entanglement characterization  
translates into  directly measurable quantities.
Out of phase oscillations of single qubit  switching probabilities 
signals two-qubit states anti-correlations~\cite{Steffen}. It follows from
$P^{({}^{1}_{2})}(t) = P(t) \pm {\rm Re}[\rho_{12}(t)]$, with 
$P(t)= - \frac{1}{2} \, \cos \varphi \,
\left [ \rho_{11}(t) + \rho_{22}(t)  \right ] + \cos^2\left (\frac{\varphi}{2}\right )$.
For a charge-phase $\sqrt{{\rm i-SWAP}}$ gate~\cite{nguyen},
 defocusing due to $1/f$ polarization 
and phase noise
is considerably reduced at optimal coupling, Figure \ref{switching}.
Phase quantum noise on qubit $2$ displaced by its optimal point
contributes to initial decay.
Oscillations visibility is larger than $90 \%$ until $\approx 10$~ns, corresponding 
to $\approx 25$ SWAP cycles. This contrasts with strong initial decay
for non-optimal coupling. 
Long-time exponential decay is due to polarization quantum noise.

In conclusion, we have proposed a general route to reduce inhomogeneities due to $1/f$ 
noise by exploiting tunability of nanodevices.
As a relevant and timely issue, we have illustrated the considerable improvement of the 
entangled dynamics in a universal two-qubit gate. 
It requires preliminary noise characterization and work in the protected SWAP-subspace. 
In the considered scheme, coupling is controllable by dynamically tuning the qubit 
frequencies.  We demonstrated that an efficient gate may be obtained with no additional
dynamical decoupling protocols, even
if at least one qubit has to be moved away from its own optimal point.
In order to implement a scalable architecture, this scheme has to be
supplemented by switchable coupling between small sub-units, formed by 
single qubit and
universal two-qubit gates~\cite{Nielsen}. Note that
our approach can also be applied to directly switchable coupling 
schemes which have been recently proposed as alternative, potentially 
scalable, designs~\cite{coupled-th-tunable,coupled-exp}.

In the broader perspective of fault tolerant architectures in the solid state, 
our work provides a strong hint on how to extend the analysis of 
Ref.~\cite{knill} to include the effect of time-correlated 
phase errors, typically affecting  nanodevices.
Note, in this respect, that fixed coupling schemes are
at the basis of qubit encoding~\cite{knill}. 

Finally, we remark that the device  reliability may  be 
qualified by the impact of one/few impurities strongly coupled to 
it~\cite{PRL05,Altshuler}. These induce non-Gaussian fluctuations 
which randomly displace qubits from 
resonant condition, possibly resulting in limited readout fidelity~\cite{nguyen}. 
These effects are neglected the present analysis, but could 
be addressed as suggested in Ref.~\cite{PRL05}.

\noindent We acknowledge discussions with D.~Vion, U. Weiss, A.~G. Maugeri and
support from EU-EuroSQIP (IST-3-015708-IP).


\begin{thebibliography}{0}
\bibitem{quantroswap} 
	Working point $q_{i,x}=1/2$, $\delta_1=0$, resonance is achieved tuning 
	$\delta_2 \approx 0.45$~\cite{nguyen}.
	Noise figures are extrapolated from single-qubit data~\cite{kn:vion}. 
	Since for both qubits  $q_{i,x}=1/2$, 
	$\Sigma_{x,i}= \Sigma_x \approx 0.02 \,\Omega$.  
	$\Sigma_{z_2} \approx 8 \times 10^{-4} \, \Omega$,
	phase noise on qubit $1$ operating at $\delta_1=0$ is negligible, 
	$\Sigma_{z_1}/\Omega \approx 10^{-5}$.
\end{thebibliography}
\end{document}